\documentclass[a4paper, 10pt]{article}
\usepackage{amsmath, amssymb, amsfonts, graphicx,amsthm}

\newtheorem{theorem}{Theorem}[section]
\newtheorem{example}{Example}[section]
\newtheorem{definition}{Definition}[section]
\renewcommand{\t}{\textrm}

\begin{document}

\title{An Optimal Multi-Unit Combinatorial Procurement Auction with Single
Minded Bidders}
\author{Sujit Gujar and Y Narahari \\
Electronic Commerce Laboratory \\Computer Science and Automation\\
Indian Institute of Science, Bangalore, 560012\\
{\tt sujit,hari@csa.iisc.ernet.in}
}


\maketitle
\begin{abstract}
The current art in optimal combinatorial auctions is limited to handling
the case of single units of multiple items, with each bidder bidding on
exactly one bundle (single minded bidders). This paper extends the current
art by proposing
an optimal auction for procuring multiple units of 
multiple items when the bidders are 
single minded. 
The auction minimizes the cost of procurement while
satisfying Bayesian incentive compatibility and interim
individual rationality. Under appropriate regularity conditions,
this optimal auction also satisfies dominant strategy
incentive compatibility. 
When bidders submit X-OR bids on two disjoint bundles, 
we show how the current state of the art can be used for 
developing optimal auctions.

\end{abstract}

\section{Introduction}
\label{sec:intro}
\subsection{Motivation and Background}
\label{ssec:motivate}
Auction based mechanisms are extremely relevant in modern
day electronic procurement systems \cite{CHANDRA07,NARAHARI09a} since they enable a promising way of
automating negotiations with suppliers and achieving the ideal
goals of procurement efficiency and cost minimization.
In many cases it may be
beneficial to allow the suppliers to bid on combinations of items rather than
on single items.  Such auctions are called
{\it combinatorial auctions}.
Simply defined, a combinatorial auction is a mechanism where bidders can submit
bids on combinations of items.  The winner determination problem is to select a
winning set of bids such that each item to be bought is included in at least one
of the selected bids, and the total cost of procurement is minimized.  
In this paper, our interest is in multi-unit combinatorial 
procurement auctions, where a buyer is interested in 
procuring multiple units of multiple items.

In mechanism design literature, an optimal auction refers to an
auction which optimizes a performance metric (for example maximizes revenue to a
seller or minimizes cost to a buyer) subject to two critical
game theoretic properties: (1) {\em incentive compatibility\/} and
(2) {\em individual rationality\/}. Incentive compatibility comes in two
forms: dominant strategy incentive compatibility (DSIC) and Bayesian
incentive compatibility (BIC). DSIC property is a property that
guarantees that reporting true valuations (or costs as the case may be)
is a best response for each bidder, irrespective of the valuations (or costs)
reported by the other bidders. BIC is a much weaker property which
ensures that truth revelation is a best response for each bidder
whenever the other bidders are also truthful.
Individual rationality (IR) is a property which assures non-negative utility
to each participant in the mechanism thus ensuring their voluntary
participation. The IR property may be (1) ex-ante IR (if the bidders
decide on participation even before knowing their exact types (valuations or costs)
or (2) interim IR (if the bidders decide on participation just after observing their
types), or ex-post IR (if the bidders can withdraw even after the game is over).
For more details on these concepts, the reader is referred to \cite{Garg08a,Garg08b,Mascollel,MYERSON81}.

\subsection{Contributions and Outline}
\label{ssec:review} 
In his seminal work, Myerson \cite{MYERSON81} characterized an optimal 
auction for selling a single unit of a single item. Extending his 
work has been attempted by several researchers and there 
have been some generalizations of his work for multi-unit single
item auctions \cite{Malakhov05,Iyengar08,Raghav07}.
Armstrong \cite{Armstrong00} characterized an optimal auction 
for two objects where type sets are binary. 
Malakhov and Vohra \cite{Malakhov05} studied an optimal 
auction for a single item multi-unit procurement auctions 
using a network interpretation. An implicit assumption in the above papers
is that the sellers have limited capacity for the item. 
They also  assume that the valuation sets are discrete. 
Kumar and Iyengar \cite{Iyengar08} and  Gautam, Hemachandra, 
Narahari, Prakash \cite{Raghav07} have proposed an optimal auction for
multi-unit, single item procurement.

Recently, Ledyard \cite{Ledyard07a} has looked at single 
unit combinatorial auctions in the presence of single minded bidders.
A \emph{single minded bidder} is one who only bids on
a particular subset of the items. 
Ledyard's auction, however, does not take into account multiple units of multiple items
and this motivates our current work which
extends Ledyard's auction to the case of procuring
multiple units of multiple items. The following are our specific contributions.
\begin{enumerate}
\item We characterize Bayesian incentive compatibility
and interim individual rationality for  procuring multiple
units of multiple items when the bidders are single minded, by deriving a 
necessary and sufficient condition.
\item  We design an optimal auction 
that minimizes the cost of procurement while
satisfying Bayesian incentive compatibility and interim
individual rationality. 
\item We show
under appropriate regularity  conditions that the proposed
optimal auction also satisfies dominant strategy
incentive compatibility.
\end{enumerate}
Some of the results presented here appeared in our paper
\cite{GUJAR09a}.

The rest of the paper is organized as follows. First, we will 
explain our model in Section \ref{sec:model} and describe the
notation that we use. We also outline certain essential technical details
of optimal auctions from the literature.
In Section 3, we present the three contributions listed above.
Section 4 concludes the paper.
\section{The Model}
\label{sec:model} 
We consider a scenario in which there is a buyer and 
multiple sellers. The buyer is interested in procuring a set 
of distinct objects, $I$. She is interested in procuring multiple 
units of each object. She specifies her demand for each object. 
The sellers are \emph{single minded}. That is each seller
is interested in selling a specific bundle of the objects. We illustrate
through an example below.

\begin{example}
Consider a buyer
interested in buying $100$ units of $A$, $150$ units of $B$,
and $200$ units of $C$. Assume that there are three sellers.
Seller 1 might be interested in providing $70$ 
units of bundle $\{A,B\}$, that is, $70$ units of $A$ and $70$ 
units of $B$ as a bundle. Because he is single minded, he does 
not bid for any other bundles. We also assume that he would supply 
equal numbers of $A$ and $B$. Similarly, seller 2 may provide
a bid for $100$ units of the bundle $\{B,C\}$. The bid from
seller 3 may be $125$ units of the bundle $\{A,C\}$.
\end{example}

The sellers are capacitated i.e. there is a maximum quantity of the
bundle of interest they could supply.
The bid therefore specifies a unit cost of the bundle and the maximum quantity that can
be supplied.
After receiving these bids, the buyer will
determine the allocation and payment as per auction rules.

We summarize below important assumptions in the model.
\label{ssec:assumptions}
\begin{itemize}
\item The sellers are single minded.
\item The sellers can collectively fulfill the demands 
specified by the buyer.
\item The sellers are capacitated i.e. they can not supply beyond the capacity 
specified in the bids.
\item The seller will never inflate his capacity, as it can 
be detected. If he fails to supply the quantity exceeding 
his capacity, he incurs a penalty which is deterrent on inflating his capacity.
This is an important assumption.
\item Whenever a buyer buys anything from the seller, she will 
procure the same number of units of each of the items from the seller's bundle 
of interest. 
\item All the participants are rational and intelligent.
\end{itemize}
Table \ref{tab:notation} shows the notation that will be 
used in the rest of the paper.
\begin{table*}
\centering
\caption{Notation}
\label{tab:notation}
\begin{normalsize}
\begin{tabular}{|l|l|}
\hline 
$I$ & Set of items the buyer is interested in buying, $\{1,2,\ldots,m\}$\tabularnewline
\hline 
$D_{j}$ & Demand for item $j$, $j=\ldots m$\tabularnewline
\hline 
$N$ & Set of sellers. $\{1,2,\dots,n\}$\tabularnewline
\hline 
$c_i$ & True cost of production of one unit of bundle of interest to the seller $i$,\tabularnewline
	& $c_i\in [ \underline{c_i},\bar{c_i} ]$\tabularnewline
\hline 
$q_i$ & True capacity for bundle which seller $i$ can supply, 
		$q_i\in[\underline{q_i},\bar{q_i}]$\tabularnewline
\hline 
$\hat{c_i}$ & Reported cost by the seller $i$\tabularnewline
\hline 
$\hat{q_i}$ & Reported capacity by the seller $i$\tabularnewline
\hline 
$\theta_{i}$ & True type i.e. cost and capacity of the seller $i$, $\theta_{i}=(c_i,q_i)$\tabularnewline
\hline 
$b_i$ & Bid of the seller $i$. $b_i=(\hat{c_i},\hat{q_i})$\tabularnewline
\hline 
$b$ & Bid vector, $(b_{1},b_{2},\ldots,b_{n})$\tabularnewline
\hline 
$b_{-i}$ & Bid vector without the seller $i$, i.e. $(b_{1},b_{2},\ldots,\, b_{i-1},b_{i+1},\ldots,\, b_{n})$\tabularnewline
\hline 
$t_{i}(b)$ & Payment to the seller $i$ when submitted bid vector is $b$\tabularnewline
\hline 
$T_i(b_i)$ & Expected payment to the seller $i$ when he submits bid $b_i$. \tabularnewline
 & Expectation is taken over all possible values of $b_{-i}$\tabularnewline
\hline 
$x_i=x_{i}(b)$ & Quantity of the bundle to be procured from the seller $i$\tabularnewline
 & when the bid vector is $b$\tabularnewline
\hline 
$X_i(b_i)$ & Expected quantity of the bundle to be procured from the seller $i$  \tabularnewline
 & when he submits bid $b_i$.\tabularnewline
 & Expectation is taken over all possible values of $b_{-i}$\tabularnewline
\hline 
$f_{i}(c_i,q_i)$ & Joint probability density function of $(c_i,q_i)$\tabularnewline
\hline 
$F_{i}(c_i,q_i)$ & Cumulative distribution function of $f_{i}(c_i,q_i)$\tabularnewline
\hline 
$f_{i}(c_i|q_i)$ & Conditional probability density function of production cost\tabularnewline
& when it is given that the capacity of the seller $i$ is $q_i$\tabularnewline
\hline 
$F_{i}(c_i|q_i)$ & Cumulative distribution function of $f_{i}(c_i|q_i)$\tabularnewline
\hline 
$H_{i}(c_i,q_i)$ & Virtual cost function for seller $i$, \tabularnewline
	&  $H_{i}(c_i,q_i)=c_i+\frac{F_{i}(c_i|q_i)}{f_{i}(c_i|q_i)}$\tabularnewline
\hline 
$\rho_{i}(b_i)$ & Expected offered surplus to seller $i$, when his bid is $b_i$\tabularnewline
\hline 
$u_{i}(b,\theta_{i})$ & Utility to seller $i$, when bid vector is $b$ and his type is
$\theta_{i}$ \tabularnewline
\hline 
$U_i(b_i,\theta_{i})$ & Expected utility to the seller $i$, when he submits bid $b_i$ and
his \tabularnewline
 & type is $\theta_{i}$. Expectation is taken over all possible values of $b_{-i}$\tabularnewline
\hline 
\end{tabular}
\end{normalsize}
\end{table*}

\subsection{Some Preliminaries}
\label{sec:opt}
The problem of designing an optimal mechanism was first studied by 
Myerson \cite{MYERSON81} and Riley and Samuelson \cite{Riley81}. 
Myerson's work is  more general and  considers the setting of 
a seller trying to sell a single unit of a single object to one of several 
possible buyers. Note here that, unlike the rest of paper, the auctioneer is the seller
and his objective is to maximize the revenue. (In the rest of the paper,
the auctioneer will be a buyer and her objective will be to minimize the cost
of procurement.) So in this particular setting, as per notation defined 
in Table \ref{tab:notation}, $m=1, D_1=1$. 
(So, $q_i$ will be 1 for all the agents and no longer a private 
information). $F_i$, $H_i$ defined in Table \ref{tab:notation} will be 
function of single variable. The buyer's private information will be the maximum 
cost he is willing to pay, which we will denote as $\theta_i$. $\theta_i \in 
\Theta_i = [\underline{\theta_i},\overline{\theta_i}]$. 

Myerson \cite{MYERSON81} characterizes all auction mechanisms
that are Bayesian incentive compatible and interim individually 
rational in this setting. From this, he derives the
allocation rule and the payment function for the optimal auction mechanism, using
an interesting notion called the
virtual cost function, defined as follows:
$$ H_i(\theta_i)= \theta_i - \frac{1-F_i(\theta_i)}{f_i(\theta_i)}$$ 
He has shown that an optimal auction is one with allocation rule as:
\begin{eqnarray}
    x_i(\theta) &=& 1 \textrm{ if }H_i(\theta_i) > \max\Big\lbrace 0, 
					\max_{j\neq i}H_j(\theta_j)\Big\rbrace \nonumber \\
		&=& 0 \textrm{ otherwise} \label{eq:opt_myerson1}
\end{eqnarray}

\begin{eqnarray}
  T_i(\theta_i) &=& E_{b_{-i}}(u_i(\theta) - \theta_i(x_i(\theta))) \nonumber\\
		&=& U_i(\theta_i) - \theta_iX_i(\theta_i) \nonumber \\
		&=& \int_{\underline{\theta_i}}^{\theta_i}X_i(s)ds - \theta_iX_i(\theta_i)
		\label{eq:opt_myerson1_pay}
\end{eqnarray}
One such payment rule is given by,
$$t_i(\theta_i,\theta_{-i}) = \Big( \int_{\underline{\theta_i}}^{\theta_i}x_i(s,\theta_{-i})ds \Big) - \Big( \theta_ix_i(\theta)\Big) \; \forall \theta$$

Any auction for single unit of an single item which satisfies 
Equation (\ref{eq:opt_myerson1}) and Equation (\ref{eq:opt_myerson1_pay}) 
is optimal i.e. maximizes seller's revenue and is BIC and IIR. 

\textit{Regularity Assumption}: If $H_i(\theta_i)$ is increasing 
with respect to $\theta_i$, then we say, the virtual cost function 
is regular or regularity condition holds true. 
Under this assumption one such optimal auction is,
\begin{enumerate}
\item Collect bids from the buyers
\item Sort them according to their virtual costs
\item If the highest virtual cost is positive, allocate 
	the object to the corresponding bidder
\item The winner, say $i$, will pay $t_i(\theta_{-i})$\\
 $= \inf \{ \theta_i | H_i(\theta_i) > 0 
	\textrm{ and } H_i(\theta_i) > H_j(\theta_j) \forall j \neq 
			i\}$
\end{enumerate}

From the payment rule, it is a dominant strategy for each bidder
to bid truthfully under the regularity assumption. 
When bidders are symmetric, i.e. $F_i$ is same 
$\forall i$, then the above optimal auction is Vickrey's second 
price auction \cite{VICKREY61}.

Myerson's work can be easily extended to the case of multi-unit auctions with 
unit demand. But problems arise when the unit-demand assumption 
is relaxed.
We move into a setting of multi-dimensional type information which 
makes truth elicitation non-trivial. Several attempts have 
addressed this problem, albeit under some restrictive assumptions
 \cite{Malakhov05,Iyengar08,Raghav07}.
It is assumed, for example, that even though the seller 
is selling multiple units (or even objects), the type information of the 
entities is still one dimensional \cite{Chen04,Dasgupta89,Zhang05}.

Researchers have also worked on extending Myerson's work for 
an optimal auction for multiple objects. The private information, 
in this setting may not be single dimensional. Armstrong 
\cite{Armstrong00} has solved this problem for two object case, 
when type sets are binary by enumerating all incentive compatibility 
conditions. Recently, Ledyard \cite{Ledyard07a} has characterized 
an optimal multi-object single unit auction, when bidders are 
single minded. 

\section{Optimal Multi-Unit Combinatorial Procurement Auction}
\label{sec:opt_auc} 
We will start this section with an example to illustrate that in a multi-unit,
multi-item procurement auction, the suppliers may have an incentive to misreport
their costs.

\begin{example}{}
Suppose the buyer has a requirement for 1000 units.
Also, suppose that there are four suppliers with $(c_i, q_i)$ values of 
$S_1 \: : \: (10, 500), \: S_2\: : \:(8, 500), \: S_3\: :\: (12, 800)$ 
and $S4\: :\: (6, 500)$. Suppose the buyer conducts the classic $k^{th}$ 
price auction, where the payment to a supplier is equal to the
cost of the first losing supplier. In this case, the sellers will be 
able to do better by misreporting types. To see this, consider that all 
suppliers truthfully bid both the cost and the quantity bids. 
The allocation then would be 
$S_1\: : \:0,\: S_2\: : \:500, \:S_3 \: : \:0, \:S_4\: :\: 500$ 
and this minimizes the total payment. Under this
allocation the payment to $S_4$ would be $10\times 500 = 5000$ 
currency units. However, if he bids his quantity
to be $490$, then the allocation changes to 
$S_1\: : \:10,\: S_2\: :\: 500,\: S_3\: :\: 0,\: S_4\: :\: 490$ 
giving him a payment of $12 \times 490 = 5880$ currency units and thus 
incentive compatibility does not hold. Thus it is evident that such 
uniform price mechanisms are not applicable to the case where both
unit cost and maximum quantity are private information. 
The intuitive explanation for this could be that by under reporting 
their capacity values, the suppliers create an artificial scarcity 
of resources in the system. 
Such fictitious shortages force the buyer to pay overboard for use 
of the virtually limited resources. 

We also make another observation 
here. Suppose, the seller $4$ bids (6,600). Then the buyer will 
order from him 600 units at the cost of 10 per unit. Being his capacity 
500, he would not be able to supply the remaining 100 units. 
If he bids (6,1000), then he will be paid only 8 per unit and 
the buyer will be ordering him 1000 units. This clearly indicates 
our assumption that a seller will not inflate his capacity is quite natural.
\end{example}

We are interested in designing an optimal mechanism, for a buyer, 
that satisfies Bayesian incentive compatibility (BIC) and 
individual rationality (IR). BIC means that the 
best response of each seller is to bid truthfully if all the 
other sellers are bidding truthfully. 
IR implies the players have non-negative 
payoff by participating in the mechanism. More 
formally, these can be stated as (see Table \ref{tab:notation} for notation), \\
$\forall i \in N$ and $\forall \;\theta_{i} \in 
	[\underline{c_i},\bar{c_i}]\times[\underline{q_i},\bar{q_i}] $
\begin{eqnarray}
U_i(\theta_{i},\theta_{i}) &\geq& 
	U_i(b_i,\theta_{i})\forall \;\;b_i,\;\;\mbox{(BIC)}\label{eq:bic}\\
U_i(\theta_{i},\theta_{i}) &\geq& 0 \qquad\qquad\qquad\; \mbox{(IR)}\label{eq:ir}
\end{eqnarray}
The IR condition above corresponds to interim individual rationality.


\subsection{Necessary and Sufficient Conditions for BIC and IR}
\label{ssec:ana}
To make the sellers report their types truthfully, the buyer has to offer 
them incentives. We propose the following incentive, motivated by paying 
a seller higher than what he claims to be the total cost of the 
production for the ordered quantity.  $\forall i \in N,$
$$ \rho_i(b_i)=T_i(b_i)-\hat{c_i}X_i(b_i), \mbox{ where } b_i=(\hat{c_i},\hat{q_i})$$
$\Rightarrow $
\begin{eqnarray}
U_i(b_i,\theta_{i}) &=& T_i(b_i) - c_i X_i(b_i) \nonumber\\
		&=& \rho_{i}(b_i) -(c_i-\hat{c_i})X_i(b_i) \label{eq:rho_utility}
\end{eqnarray}
With the above offered incentive, we now state and prove the following theorem.
\begin{theorem}
\label{thm:bic_ir}
Any mechanism in the presence of single minded, capacitated sellers is BIC and IR iff
	\begin{enumerate}
	\item $\rho_{i}(b_i) = \rho_{i}(\bar{c_i,}\hat{q_i}) + 	
		\int_{\hat{c_i}}^{\bar{c_i}}X_i(t,\hat{q_i})dt$

	\item $\rho_{i}(b_i)$ non-negative, and non-decreasing in 
		$\hat{q_i}\;\;\forall\;\hat{c_i}\;\in\;[\underline{c_i},\bar{c_i}]$

	\item The quantity which seller $i$ is asked to supply, $X_i(c_i,q_i)$
	is non-increasing in $c_i\;\;\forall q_i\;\in\;[\underline{q_i},\bar{q_i}]$.
	\end{enumerate}
\end{theorem}
\vspace*{5mm}
\noindent\begin{proof}: A similar
theorem is presented by Kumar and Iyengar \cite{Iyengar08}
for the case of multi-unit single item procurement auctions. 
Using the notion of single minded bidder \cite{Ledyard07a}, 
we state and prove a  result for a wider setting.

To prove the necessity part of the theorem, we first observe that,
$$U_i(b_i,\theta_{i}) = U_i(\hat{c_i},\hat{q_i},c_i,q_i) 
= T_i(b_i)-c_iX_i(b_i) $$
$$\mbox{and BIC }\Rightarrow U_i(\hat{c_i},\hat{q_i},c_i,q_i) \leq  U_i(c_i,q_i,c_i,q_i), $$ 
$$\forall(\hat{c_i},\hat{q_i}) \mbox{ and }(c_i,q_i)\in \Theta_i$$
In particular, 
$$U_i(\hat{c_i},q_i,c_i,q_i)\leq U_i(c_i,q_i,c_i,q_i)$$
Without loss of generality, we assume $\hat{c_i}>c_i.$ 
Rearrangement of these terms yields,
$$U_i(\hat{c_i},q_i,c_i,q_i) = U_i(\hat{c_i},q_i,\hat{c_i},q_i) $$
$$\quad\qquad+ (\hat{c_i}-c_i)X_i(\hat{c_i},q_i)$$
$\Rightarrow$ 
$$\frac{U_i(\hat{c_i},q_i,\hat{c_i},q_i)-U_i(c_i,q_i,c_i,q_i)}{\hat{c_i}-c_i}
 \leq -X_i(\hat{c_i},q_i)$$
Similarly using, $$U_i(c_i,q_i,\hat{c_i},q_i) \leq U_i(\hat{c_i},q_i,\hat{c_i},q_i)$$ 
\begin{equation}
-X_i(c,q)\leq\frac{U_i(\hat{c_i},q_i,\hat{c_i},q_i)-U_i(c_i,q_i,c_i,q_i)}{\hat{c_i}-c_i}\nonumber
\end{equation}
\begin{equation} \leq-X_i(\hat{c_i},q_i).\label{eq:mono}
\end{equation}

Taking limit $\hat{c_i}\rightarrow c_i,$ we get,
\begin{equation}
\frac{\partial U_i(c_i,q_i,c_i,q_i)}{\partial{c_i}} = -X_i(c_i,q_i). 
\label{eq:pde}
\end{equation}
Equation (\ref{eq:mono}) implies, $X_i(c_i,q_i)$ is non-increasing in $c_i$. 
This proves statement 3 of the theorem in the forward direction.
When the seller bids truthfully, from Equation (\ref{eq:rho_utility}),
\begin{equation}
\rho_{i}(c_i,q_i)=U_i(c_i,q_i,c_i,q_i).\label{eq:rho1}
\end{equation}
For BIC, Equation (\ref{eq:pde}) should be true. So,
\begin{equation}
\rho_{i}(c_i,q_i)=\rho_{i}(\bar{c_i},q_i)+\int_{c_i}^{\bar{c_i}}X_i(t,q_i)dt\label{eq:rho2}
\end{equation}
This proves claim 1 of the theorem. BIC also requires,
$$q_i \in \arg\max_{\hat{q_i}\in[\underline{q_i},q_i]} U_i(c_i,\hat{q_i},c_i,q_i)
\;\forall\; c_i\;\in\;[\underline{c_i},\bar{c_i}]$$

(Note that $\hat{q_i}\in [\underline{q_i},q_i] \mbox{ and not } \in[\underline{q_i},\bar{q_i}]$ 
as it is assumed that a bidder will not over report his capacity.) 

This implies, $\forall\; c_i\;\;\rho_{i}(c_i,q_i)$ should be non-decreasing in 
$q_i.$. The
IR conditions (Equations (\ref{eq:ir}) and (\ref{eq:rho1})) imply
$$\rho_{i}(c_i,q_i)\geq 0.$$
This proves statement 2 of the theorem. Thus, these three 
conditions are necessary for BIC and IR properties.

We now prove that these are sufficient conditions for BIC and IR.
Assume that all three conditions are true,
$$\Rightarrow U_i(\theta_{i},\theta_{i})=\rho_i(c_i,q_i) \geq 0.$$
So the IR property is satisfied.
\begin{eqnarray}
U_i(b_i,\theta_{i}) &=& 
	\rho_{i}(\hat{c_i},\hat{q_i})+(\hat{c_i}-c_i)X_i(\hat{c_i},\hat{q_i})\nonumber \\
 &=& \rho_{i}(\bar{c_i},\hat{q_i}) + \int_{\hat{c_i}}^{\bar{c_i}}X_i(t,\hat{q_i})dt\nonumber \\ 
	& & \quad + (\hat{c_i}-c_i)X_i(\hat{c_i},\hat{q_i}) \nonumber \\
 &=& \rho_{i}(\bar{c_i},\hat{q_i}) + \int_{c_i}^{\bar{c_i}}X_i(t,\hat{q_i})dt\nonumber \\ 
	& & \quad - \int_{c_i}^{\hat{c_i}}X_i(t,\hat{q_i})dt \nonumber \\ 
	& & \quad+ (\hat{c_i}-c_i)X_i(\hat{c_i},\hat{q_i})\nonumber \\
 &\leq& \rho_{i}(c_i,\hat{q_i})\nonumber \\ 
	& & \mbox{ as }X_i\mbox{ is non-increasing in }c_i \nonumber \\
 &\leq& \rho_{i}(c_i,q_i) \nonumber \\
 &=& U_i(\theta_{i},\theta_{i}) \:\;\nonumber \\ 
	& & \mbox{ as }\rho_{i}\mbox{ is non-decreasing in }q_i \nonumber
\end{eqnarray}
This proves the sufficiency of the three conditions.\\
\end{proof}

\subsection{Allocation and Payment Rules of Optimal Auction}
\label{ssec:oa}
The buyer's problem is to solve,
\begin{center}
$\min\:\mathbf{E}_{b} \sum_{i=1}^{n} t_{i}(b) \quad \mbox{s.t.}$                                                                \end{center}
\begin{enumerate}
\item $t_{i}(b) = \rho_{i}(b) + \hat{c_i}x_{i}(b)$
\item All three conditions in Theorem \ref{thm:bic_ir} hold true. 
\item She procures at least $D_{j}$ units of each item $j$.
\end{enumerate}

Expectation being a linear operator, the buyer's problem is 
to minimize $\sum_{i=1}^{n} {\mathbf{E}}_{b_i} T_i(\hat{c_i},\hat{q_i})$.
Condition 1 of the theorem has to hold true, which will
imply the $i^{th}$ term in the summation is given by,
\begin{center}
$\int_{\underline{q_i}}^{\bar{q_i}} \int_{\underline{c_i}}^{\bar{c_i}} 
\Big(c_iX_i(c_i,q_i) + \rho_{i}(\bar{c_i},q_i)\qquad $\\$\quad\quad+ 
\int_{c_i}^{\bar{c_i}}X_i(t,q_i)dt\Big)f_{i}(c_i,q_i)dc_idq_i$
\end{center}
However,
\begin{center}
$\int_{\underline{c_i}}^{\bar{c_i}} \left(\int_{c_i}^{\bar{c_i}}
X_i(t,q_i)dt\right)f_{i}(c_i,q_i)dc_i =\qquad$\\$\qquad \int_{\underline{c_i}}^{\bar{c_i}}
X_i(c_i,q_i)F_{i}(c_i|q_i)f_{i}(q_i)dc_i$\end{center}

Condition 2 of Theorem \ref{thm:bic_ir} requires 
$\rho_{i}(\bar{c_i},q_i) \geq 0 $ and the buyer wants to minimize 
the total payment to be made. So, she has to assign $\rho_{i}(\bar{c_i},q_i) = 0 
\; \forall\; q_i ,\forall i$. So her problem is to solve,
\begin{center}
$\min\,\sum_{i=1}^{n} \int_{\underline{q_i}}^{\bar{q_i}} \int_{\underline{c_i}}^{\bar{c_i}} \left(c_i+\frac{F_{i}(c_i|q_i)}{f_{i}(c_i|q_i)}\right) \qquad\qquad$
\\ $\qquad\qquad\qquad X_i(c_i,q_i)f_{i}(c_i,q_i)dc_idq_i $
\end{center}
That is,
\begin{center}
$\min\,\sum_{i=1}^{n} \int_{\underline{q_i}}^{\bar{q_i}} \int_{\underline{c_i}}^{\bar{c_i}} H_{i}(c_i,q_i)X_i(c_i,q_i)\qquad$\\
$\qquad\qquad f_{i}(c_i,q_i)dc_idq_i$
\end{center}
where, $H_{i}(c_i,q_i)$ is the virtual cost function, defined in 
Table \ref{tab:notation}. Define,
\begin{center}
	$\bar{c} = (\bar{c_{1}},\bar{c_{2}},\ldots,\bar{c_{n}})$ \\
	$c = (c_{1},c_{2},\ldots,c_{n})$ \\
	$\underline{c} = 
		(\underline{c_{1}},\underline{c_{2}},\ldots,\underline{c_{n}})$.
\end{center}
Similarly, define $\bar{q}$ , $q$ and $\underline{q}.$ Let, 
\begin{center}
	$dc = dc_{1}dc_{2}\ldots dc_{n}$\\ 
	$dq = dq_{1}dq_{2}\ldots dq_{n}$ \\
	$f(c,q) = \prod_{i=1}^{n}f_{i}(c_i,q_i) $
\end{center}
Her problem now reduces to,
\begin{center}
	$\min \int_{\underline{q}}^{\bar{q}} \int_{\underline{c}}^{\bar{c}} 
			\left(\sum_{i=1}^{n}H_{i}(c_i,q_i)x_{i}(c_i,q_i)\right)$\\
	$\qquad\qquad f(c,q)dcdq\quad\mbox{s.t.}$
\end{center}
1. $\forall\: i,\; X_i(c_i,q_i)$ is non-increasing in $c_i, \forall\: q_i. $\\
2. The Buyer's minimum requirement of each item is satisfied. \\
This is \emph{an optimal auction} for the buyer in the presence of 
the single minded sellers.

In the next subsection, we will see an optimal auction under 
regularity conditions.
\subsection{Optimal Auction under Regularity Assumption}
\label{ssec:regular}
First, we make the assumption that,
$$H_{i}(c_i,q_i) = c_i + \frac{F_{i}(c_i|q_i)}{f_{i}(c_i|q_i)}$$
is non-increasing in $q_i$ and non-decreasing in $c_i$. This 
regularity assumption is 
the same as regularity assumption made by Kumar and Iyengar 
\cite{Iyengar08}.
With this assumption, the buyer's optimal 
auction when bidder $i$ submits bid as $(c_i,q_i)$ is,
$$\min\sum_{i=1}^{n} x_{i}H_{i}(c_i,q_i)\quad\quad\mbox{subject to}$$
\begin{enumerate}
\item $0 \leq x_{i} \leq q_i$, where $x_{i}$ denotes the quantity 
that seller $i$ has to supply of bundle $\bar{x_i}$.
\item Buyer's demands are satisfied.
\end{enumerate}

The condition $X_i(c_i,q_i)$ is non
increasing in $c_i, \forall \:q_i$ and $\forall \:i$.
After this problem has been solved, the buyer pays each seller 
$i$ the amount 
\begin{equation}
    t_{i} = c_ix^*_{i} + \int_{c_i}^{\bar{c_i}}x_{i}(t,q_i)dt \label{eq:payment}
\end{equation} 
where $x^*_i$ is what agent $i$ has to supply after solving the 
above problem.

We exemplify the optimal mechanism with one example.
\begin{example}
Suppose, the buyer is interested in buying 100 units
of $\{ A,C,D\}$ and $250$ units of $\{B\}$. Seller 1 ($S1$) is interested 
in providing $q_1=100$ units of bundle $\{A,B\}$, seller 2 ($S2$): $q_2=100$ units
of $\{B\}$, seller 3, ($S3$) $q_3=150$ units of $\{B,C,D\}$ and 
seller 4 ($S4$) is interested in up to $q_4=120$ units of $\{A,B,C,D\}$.
The unit costs of the respective bundles are 
$c_1=100$, $c_2=50$, $c_3=70$ and $c_4=110$. 
Each seller will submit his bid as $(c_i,q_i)$.
After receiving the bids, buyer will solve,
$$\min x_1H_1(100,100) + x_2H_2(50,100)\qquad$$ 
$$\qquad+x_3H_3(70,150)+x_4H_4(110,120)$$
s.t.
\begin{eqnarray}
x_i &\geq& 0 \;\; i=1,2,3,4. \nonumber \\
x_1 &\leq&  100 \nonumber \\
x_2 &\leq&  100 \nonumber \\
x_3 &\leq&  150 \nonumber \\
x_4 &\leq&  120 \nonumber \\
x_1 + x_2 &\geq& 100 \label{eq:1} \\
x_1+x_2+x_3+x_4 &\geq& 250  \label{eq:2} \\
x_3+x_4 &\geq& 100 \label{eq:3}
\end{eqnarray}
Equation (\ref{eq:1}) is required to be satisfied as at least 
100 units of $A$ has to be procured. Equation (\ref{eq:2}) 
is for procuring at least 250 units of $B$, and Equation (\ref{eq:3}) 
is for procuring at least 100 units of $C$ and $D$.
After solving this optimization problem, she will determine the payment 
according to Equation (\ref{eq:payment}).
\end{example}

It can be seen that for the seller $i$, the best response is
to bid truthfully irrespective of whatever the others are bidding. 
Thus, this mechanism enjoys the stronger property, namely dominant
strategy incentive compatibility. Note that this property is much 
stronger than BIC.
The above property is a direct consequence of the result proved by 
Mookherjee, and Stefan \cite{Mookherjee92}. They
have given the monotonicity conditions for DSIC implementation 
of a BIC mechanism. Under these regularity assumptions, $x_i$ satisfies 
these conditions. So we have a DSIC mechanism.
In the next section
we consider X-OR bidding with unit demand case.

\section{An Optimal Auction when Bidders are XOR Minded}
\label{sec:xor}
Consider the situation where a supplier can manufacture some
of the items required by the buyer, say $A,B,C,D$. However,
with the machinery he has, at a time either he can manufacture
$A,D$ or $B,C$ but not any other combination simultaneously.
Thus he can either supply $A,D$ as bundle or $B,C$ as a bundle
but not both. That is, he is interested in X-OR bidding. 
\begin{definition}[XOR Minded Bidder]
We say a bidder is an {\em XOR minded} if he is interested in
supplying either of two disjoint subsets of items auctioned
for but not both.
\end{definition}

To simplify the analysis, in this section, we restrict 
ourselves to the unit demand case.
That is the buyer is interested in buying single unit of each 
of the items from $I$. And hence there are no capacity 
constraints. We formally state assumptions.
\begin{itemize}
	\item We assume that the bidders are XOR minded.
	\item  For each bidder, his costs of 
				the two bundles of his interest are independent. 
	\item The two bundles for which each seller is going
				to submit an X-OR bid, are known.
	\item The sellers can collectively supply the items 
			required by the buyer.
	\item The buyer and the sellers are strategic.
	\item Free disposal. That is, if the buyer procures more than 
			one unit of an item, he can freely dispose it of.
\end{itemize}

With the above assumptions, we now discuss an 
extension of the current art of designing optimal auctions
for combinatorial auctions in the presence of XOR 
minded bidders. Though we assume the bidders are XOR 
minded, the BIC characterization and the auction designed 
here work even though the bidders are either single minded or 
XOR minded. 

\subsection{Notation}
As, $q_i=1$ for each bidder, we drop capacity from 
the types and bids for all the agents.
Each agent will be reporting the costs for each bundle
of his interest, he will be bidding two real numbers.
And we need to calculate virtual costs on both the bundles.
Thus, we need appropriate modifications in some of the notation
used in the paper. We summarize the new notation for this
section in Table \ref{tab:notation2}. Each agent is submitting
tow different bids on two different bundles. We will use $j$ to refer to the bundle.

\begin{table*}[!hbt]
\centering
\begin{normalsize}
\caption{Notation: XOR Minded Bidders}
\label{tab:notation2}
\begin{tabular}{|l|l|}
\hline 
$j$ & $j=1\mbox{ or }2$. Bundle index.\\
$B_{i_j}$ & The $j^{th}$ bundle of items for which the agent $i$ 
			is bidding. $j=1,2$\\
\hline
$c_{i_j}$	& True cost of production of $B_{i_j}$ 
			to the seller $i$. 
			$c_{i_j}\in [ \underline{c_i},\bar{c_i} ]$\\
$c_i$ & $=(c_{i_1},c_{i_2})$ \\
\hline 
$\theta_{i}$ & True type i.e. costs for $i$, $\theta_{i}=(c_{i_1},c_{i_2})$\\
\hline 
$b_i$ & Bid of the seller $i$. 	
		$b_i=(\hat{c_{i_1}},\hat{c_{i_2}})$\\
\hline 
$x_{i_j}=x_{i_j}(b)$ & Indicator variable to indicate whether 
		$B_{i_1}$ is to be procured from \\
 & the seller $i$ when the bid vector is $b$\tabularnewline
\hline 
$X_{i_j}(b_{i})$ & Probability that $B_{i_j}$ is procured
		from the seller $i$  when he \\
	 & submits bid $b_i$. Expectation is taken over all possible values of $b_{-i}$\\
\hline 
$f_{i_j}(c_{i_j})$ &  Probability density function of $(c_{i_j})$\tabularnewline
\hline 
$F_{i_j}(c_{i_j})$ & Cumulative distribution function 
		of $c_{i_j}$\\
\hline

$H_{i_j}(c_{i_j})$ & Virtual cost function for seller $i$,
	for bundle $B_{i_j}$ \\
	&  $H_{i_j}(c_{i_j})=c_{i_j}+\frac{F_{i_j}(c_{i_j})}{f_{i_j}(c_{i_j})}$
		\\
\hline 
\end{tabular}
\end{normalsize}
\end{table*}

\subsection{Optimal Auctions When Bidders Are XOR Minded}
First we characterize the BIC and IIR mechanisms for the 
settings under consideration in next subsection.
We design an optimal auction in subsection \ref{sssec:ocax}.

\subsubsection{BIC and IIR: Necessary and Sufficient Conditions}
The utility for the agent $i$ is
$$U_i(b_i,\theta_i) = -c_{i_1}X_{i_1} - c_{i_2}X_{i_2} + 
T_i(b_i,\theta_i)$$ Using similar arguments as in
the proof of the Theorem \ref{thm:bic_ir}, for any mechanism 
in the presence of XOR minded bidders, the necessary condition
for BIC is,
\begin{eqnarray}
\frac{\partial U(.)}{\partial c_{i_1}} = X_{i_1}(c_{i_1},c_{i_2}) \nonumber\\
\frac{\partial U(.)}{\partial c_{i_2}} = X_{i_2}(c_{i_1},c_{i_2})
\label{eqn:pde}
\end{eqnarray}

\noindent and $X_{i_j}(c_{i_1},c_{i_2})$ should be non-increasing in $c_{i_j},\;j=1,2$.

\noindent We make an assumption that, 
\begin{equation}
\frac{\partial X_{i_1}}{\partial c_{i_2}} = \frac{\partial X_{i_2}}{\partial c_{i_1}}
\label{eqn:assumption}
\end{equation}

In general, the above assumption is not necessary for the mechanism to be truthful. However, if we assume that 
Equation (\ref{eqn:assumption}) is true,
we can solve PDE (\ref{eqn:pde}) analytically.
Now we can state the following theorem, 
\begin{theorem}
With assumption (\ref{eqn:assumption}), a necessary 
and sufficient condition for a mechanism to be BIC and IIR
in the presence of XOR minded bidders is,
\begin{enumerate}
	\item $T_i(.) = c_{i_1}X_{i_1} + c_{i_2}X_{i_2} 
	+\int_{(c_{i_1},c_{i_2})}^{(\overline{c_i},\overline{c_i})}
			\triangledown U_i(.)d\theta_i$
	\item $U_i(\overline{c_i},\overline{c_i}) \geq 0$.
\end{enumerate}
\end{theorem}

\subsubsection{Optimal Auction with Regularity Assumption}
\label{sssec:ocax}
Suppose we assume that, $H_{i_j}$ is non-decreasing
in $c_{i_j}$ for each $i,j$. This is the same
regularity assumption as Myerson \cite{MYERSON81}.
Now, following similar treatment for buyers problem as in Section \ref{ssec:regular},
reduces the buyers problem to:
\begin{equation}
\begin{array}{|l|}
\hline
\min\sum_{i=1}^{n}\sum_{j=1}^{2} x_{i_j}H_{i_j}(c_{i_j})\\
\mbox{subject to}\\

\t{ 1. } x_{i_j} \in {0,1}, \t{ where } x_{i_j} \t{ indicates 
		whether supplier $i$ is supplying his}s \\
		j^{th}  \t{ bundle or no.} \\
\t{ 2. } x_{i_1} + x_{i_2} \leq 1. \t{ (XOR minded bidder).}\\
\t{ 3. All the items are procured.}\\
\hline
\end{array}
\label{eqn:regular_ocax}
\end{equation}

\begin{figure}[!htb]
\centering
\includegraphics[width=13cm]{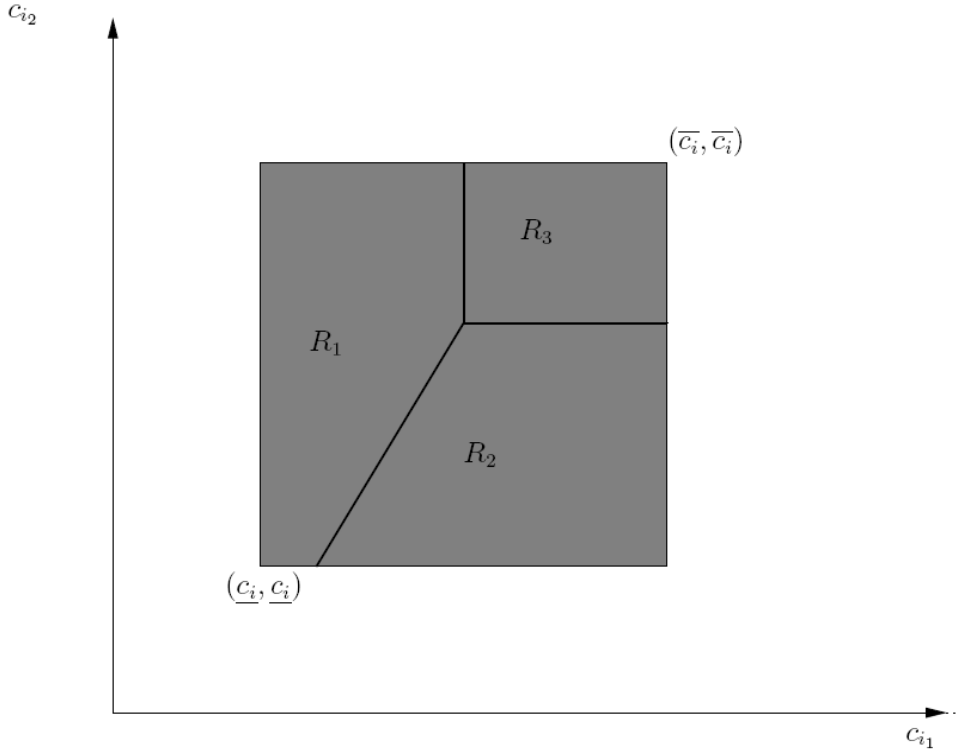}
\caption{X-OR Bidding}
\label{fig:xor}
\end{figure}

Now, we show that at optimal allocation, the assumption 
(\ref{eqn:assumption}) holds true. For an agent $i$, 
fix, $\theta_{-i}$ and consider the square of his types 
$[\underline{c_i},\overline{c_i}]\times[\underline{c_i},\overline{c_i}]$.
When he bids, $b_i=(\overline{c_i},\overline{c_i})$, he does 
not win any item. However, if he decreases his bid on 
$c_{i_j}$, he wins the bundle $B_{i_j}$ at some lower 
bid and at a lower bid for $B_{i_j}$, he continues to win.
Also, he being XOR minded, he cannot win both the bundles.
Thus, the type set's square can be partitioned into
three regions, $R_1,R_2$ and $R_3$ as shown 
in Figure \ref{fig:xor}. When his type
is in region $R_j$, he is asked to supply 
$B_{i_j}, \;j=1,2$ and when it is in $R_3$ he is not in 
the list of winning agents. Now, except on the boundary between
$R_1$ and $R_2$, the assumption (\ref{eqn:assumption})
holds true. Hence, though we are not using (\ref{eqn:assumption})
as a necessary condition, it is getting satisfied in
optimization problem (\ref{eqn:regular_ocax}). Thus
OCAX is an optimal combinatorial auction for
the buyer in the presence of XOR minded bidders.

\subsubsection{The Case when Regularity Assumption is not Satisfied}
Though we do not solve the buyer's problem of optimal 
mechanism design without the regularity assumption, we highlight
some thoughts on this. If we can assume (\ref{eqn:assumption}),
then we can design an optimal auction very similar to 
the OCAS, in the presence of XOR minded bidders.  
The challenge is, we cannot use (\ref{eqn:assumption}) 
as a necessary condition nor can we assume it. 
However, it may happen that in an optimal auction,
the condition (\ref{eqn:assumption}), will hold true. 
We are still working on this. 

\section{Conclusion}
\label{sec:conclusion}
In this paper,
\begin{itemize}
	\item we have stated and proved a necessary 
			and sufficient condition for 
			incentive compatible and individually 
			rational multi-unit multi-item auctions in the
			presence of single minded, capacitated buyers. 
	\item We have given a blueprint of an optimal 
			mechanism, for a buyer seeking to 
			procure multiple units of multiple items in
			the presence of single minded and capacitated
			sellers.
	\item We also have shown that the 
			mechanism minimizes the cost subject to DSIC and 
			IIR if the virtual cost functions satisfy the
			regularity assumptions.
	\item When bidders are XOR minded, under certain regularity
			conditions, we designed an optimal auction
			for the buyer which we call as OCAX.
\end{itemize}

There are many natural extensions to this work. 
First, we can study optimal auctions in which the 
sellers are willing to give volume discounts. 
We also plan to study a case where the 
sellers are interested in supplying multiple bundles. 

\section*{Acknowledgment}
The first author would like to acknowledge the Infosys 
Technologies Pvt Ltd for awarding Infosys fellowship to 
pursue PhD.

\bibliographystyle{acm}
\bibliography{orji}

\end{document}